\documentstyle[12pt]{article}
\begin{document}
\topmargin=-1.0cm
\evensidemargin=0cm
\oddsidemargin=0cm

%%%%%%%%%%%%%%%%%%%%%%%%%%%%%%%%%%%
%%%%%%%%%%%%%%%%%%%%%%%%%%%%%%%%%%%
%%%%%%%%%        newcommand                  %%%%%%%%%%%
%%%%%%%%%%%%%%%%%%%%%%%%%%%%%%%%%%%
%%%%%%%%%%%%%%%%%%%%%%%%%%%%%%%%%%%

\newcommand{\tilphi}{\tilde{\Phi}}

\newcommand{\bea}[1]{\begin{eqnarray} #1 \end{eqnarray}}

\newcommand{\deri}[2]{\frac{\partial #1}{\partial #2}}

\newcommand{\bra}[1]{\bigl\langle #1 \big|}

\newcommand{\ket}[1]{\big| #1 \bigr\rangle}

\newcommand{\bket}[2]{\bigl\langle #1 \big| #2 \bigr\rangle}

\newcommand{\intinf}{\int^{\infty}_{-\infty}}

\newcommand{\dpi}[2]{\frac{d^#1 #2}{(2 \pi)^#1}}

\newcommand{\spro}[2]{\vec{#1}\cdot\vec{#2}}

\newcommand{\kakko}[1]{\left(#1\right)}

\newcommand{\tkakko}[1]{\left\{#1\right\}}

\newcommand{\dkakko}[1]{\left[#1\right]}

\newcommand{\bee}[1]{\begin{equation}#1\end{equation}}

\newcommand{\lag}{{\cal L}}

\newcommand{\expect}[1]{\left\langle #1 \right\rangle}

\newcommand{\nonum}{\nonumber\\}

\newcommand{\ry}{{\rm Ry}}

\newcommand{\brkt}[3]
{\left\langle #1 \left| #2 \right| #3 \right\rangle}

\newcommand{\e}{{\rm e}}

\newcommand{\Tr}{{\rm Tr}}
%%%%%%%%%%%%%%%%%%%%%%%%%%%%%%%%%%%
%%%%%%%%%%%%%%%%%%%%%%%%%%%%%%%%%%%
%%%%%%%%%%%%%%%%%%%%%%%%%%%%%%%%%%%
%%%%%%%%%%%%%%%%%%%%%%%%%%%%%%%%%%%

\begin{titlepage}
%%%%
\begin{flushright}
{ICRR-REPORT-466-2000-10}\\
{KOBE-TH-00-06}
\end{flushright}
%%%
\vspace{1cm}
%%%%
\begin{center}
{\Large SYMMETRY BREAKING/RESTORATION }\\
{\Large IN A NON-SIMPLY CONNECTED SPACE-TIME
\footnote{Talk given at XXXth International Conference on High Energy Physics 
    (ICHEP2000), July 27-August 2, Osaka, Japan. To be published in the
    Proceedings (World Scientific, Singapore).
}}\\
\vskip1.0truein
{\large Hisaki Hatanaka
\footnote{E-mail: {\tt hatanaka@icrr.u-tokyo.ac.jp}}
}\\
\vspace*{2mm}
{\it Institute for Cosmic Ray Research,\\
University of Tokyo, Kashiwanoha 5-1-5, Kashiwa 277-8582, Japan\\
\vspace*{6mm}
 {\large Seiho Matsumoto
 \footnote{E-mail: {\tt matsumoto@oct.phys.sci.kobe-u.ac.jp}}}
{\rm and} {} {\large Katsuhiko Ohnishi
\footnote{E-mail: {\tt katuhiko@oct.phys.sci.kobe-u.ac.jp}}}\\
\vspace*{2mm}
 {\it Graduate School of Science and Technology,\\
Kobe University, Rokkodai, Nada, Kobe 657-8501, Japan}\\
\vspace*{6mm}
{\large Makoto Sakamoto
\footnote{E-mail: {\tt sakamoto@phys.sci.kobe-u.ac.jp}}}\\
\vspace*{2mm}
Department of Physics,\\
Kobe University, Rokkodai, Nada, Kobe 657-8501, Japan }
%%%%%%%%%%%%%%%%%%%%%%%%%%%%%%%%%%%%%%%%%%%%%
%%%%%%%%%%%%%%%%%%%%%%%%%%%%%%%%%%%%%%%%%%%%%
%%%%%%%%%%%%%%%%%%%%%%%%%%%%%%%%%%%%%%%%%%%%%
\vskip0.5truein

\end{center}

\vskip0.5truein \centerline{\bf Abstract} \vskip0.13truein
%%%%%%%%%%%%%%%%%%%%%%%%%%%%%%%%%%%%%%%%%%%%%
%%%%%%%%%%%%%%%%%%%%%%%%%%%%%%%%%%%%%%%%%%%%%
%%%%%%%%%%%%%%%%%%%%%%%%%%%%%%%%%%%%%%%%%%%%%
Field theories compactified on non-simply connected spaces,
which in general allow to impose twisted boundary conditions,
are found to unexpectedly have a rich phase structure.
One of characteristic features of such theories is the appearance
of critical radii,
at which some of symmetries are broken/restored.
A phase transition can occur at the classical level,
or can be caused by quantum effects.
The spontaneous breakdown of the translational invariance
of compactified spaces is another characteristic feature.
As an illustrative example,
the $O(N)$ $\phi^4$ model on $M^3\otimes S^1$ is
studied and the novel phase structure is revealed.
%%%%%%%%%%%%%%%%%%%%%%%%%%%%%%%%%%%%%%%%%%%%%
%%%%%%%%%%%%%%%%%%%%%%%%%%%%%%%%%%%%%%%%%%%%%
%%%%%%%%%%%%%%%%%%%%%%%%%%%%%%%%%%%%%%%%%%%%%

\end{titlepage}
\newpage
\baselineskip 20 pt
\pagestyle{plain}
\vskip0.2truein
\setcounter{equation}{0}
\vskip0.2truein
%%%%%%%%%%%%%%%%%%%%%%%%%%%%%%%%%%%%%%%%%%%%%
%%%%%%%%%%%%%%%%%%%%%%%%%%%%%%%%%%%%%%%%%%%%%
%%%%%%%%%%%%%%%%%%%%%%%%%%%%%%%%%%%%%%%%%%%%%
\renewcommand{\thefootnote}{\fnsymbol{footnote}}

The parameter space of field theories compactified on non-simply
connected spaces is,
in general,
wider than that of ordinary field theories on the Minkowski space-time,
and is spanned by twist parameters specifying boundary 
conditions\cite{Isham,Hosotani},
in addition to parameters appearing in the actions.
Physical consequences caused by twisted boundary conditions
turn out to be unexpectedly rich and many of them have not
been uncovered yet.
The purpose of this talk is to report some of interesting properties
of such theories overlooked so far.

One of characteristic features of such theories is 
the appearance of critical radii of compactified spaces,
at which some of symmetries are broken/restored\cite{o(n)}.
Symmetry breaking patterns are found to be unconventional.
A phase transition can occur at the classical level,
or can be caused by quantum effects.
Radiative corrections would become important when
a compactified scale is less than the inverse of a typical
mass scale, and then some of broken symmetries could be restored,
or conversely some of symmetries could be broken.
Another characteristic  and probably surprising feature is
the spontaneous breakdown of the translational invariance
of compactified spaces\cite{translation}.
Twisted boundary conditions do not allow vacuum expectation values
of twisted bosons to be non-vanishing constants.
In other words, vacuum expectation values of twisted bosons
have to vanish or to be coordinate-{\it dependent} if they are
non-vanishing.
If the minimum of a potential does not lie at the origin,
twisted bosons could acquire non-vanishing vacuum expectation
values, which should be coordinate-dependent.
Then, we have to minimize the total
energy, which consists of both the kinetic term and the
potential term, to find the vacuum configuration.
When non-vanishing vacuum expectation values of twisted
bosons are energetically preferable,
they should be coordinate-dependent and hence the translational
invariance is broken spontaneously.
Among other characteristic features, a phenomenologically
important observation is that twisted boundary conditions
can break supersymmetry spontaneously\cite{susy}\footnote{
This subject was discussed in the talk given by M. Tachibana
at this Conference.}.
This will give a new type of spontaneous supersymmetry
breaking mechanisms and it would be of great interest
to investigate a possibility to construct realistic
supersymmetric models with this supersymmetry breaking
mechanism, though this subject will not be treated
in this talk.

As an illustrative example,
we here concentrate on the $O(N)$ $\phi^4$ model
on $M^3\otimes S^1$\footnote{
The classical analysis of 
this model has been done in Ref.[3].}.
The action which consists of $N$ real scalar fields 
$\phi_i$ $ (i=1,\cdots,N)$
is given by
\bee{
        S
=
        \int d^3x\int^{2\pi R}_0dy
        \dkakko{
                \frac{1}{2}\partial_A\phi_i\partial^A\phi_i
                -
                \frac{m^2}{2}\phi_i^2
                -
                \frac{\lambda}{8}\kakko{\phi_i^2}^2
        }
\label{1}
,}
where $y$ and $R$ denote the coordinate and the radius of $S^1$,
respectively.
Since $S^1$ is multiply-connected,
we can impose a twisted boundary condition on $\phi_i$ such as
\bee{
        \phi_i(y+2\pi R)
=
        U_{ij}\phi_j(y)
.
\label{2}}
The matrix $U$ must belong to $O(N)$,
otherwise the action would not be single-valued.
We shall below consider various boundary conditions
and discuss physical consequences.
\section*{(1) $U={\bf 1}$}

In this case,
the fields $\phi_i(y)$ obey the periodic boundary condition.
For $m^2>0$,
the phase structure is trivial:
The $O(N)$ symmetry is unbroken in a whole range of $R$.
For $m^2<0$,
$O(N)$ would be broken to $O(N-1)$.
It is well known that the leading correction to the squared mass
is proportional to $1/R^2$ for small radius $R$
\cite{radiative1,radiative2} and
that the broken symmetry $O(N-1)$ can be 
restored for $R\leq R^\ast={\cal O}(\sqrt{\lambda}/\mu)$ 
$(\mu^2\equiv-m^2)$,
just like the symmetry restoration at high
temperature\cite{temperature}.
Thus, we have found no new interesting phenomena
with $U={\bf 1}$.
\section*{(2) $U=-{\bf 1}$}

In this case,
$\phi_i(y)$ obey the antiperiodic boundary condition.
For $m^2>0$,
nothing happens and the $O(N)$ symmetry remains unbroken in a whole 
range of $R$,
while for $m^2<0$,
several interesting phenomena occur\cite{o(n)}.
For $R>R^\ast\sim 1/(2\mu)$,
the $O(N)$ symmetry is spontaneously broken to $O(N-2)$
but {\it not} $O(N-1)$!
The translational invariance of $S^1$ is also broken spontaneously
because of the $y$-dependent vacuum expectation values of $\phi_i(y)$.
For $R\leq R^\ast$, all the broken symmetries are restored.
It should be emphasized that the mechanism of this symmetry
restoration is different from the previous case of $U={\bf 1}$
and that the present symmetry restoration has a classical origin. 
This may be seen from the fact
that $R^\ast$ is of order $1/\mu$ but not $\sqrt{\lambda}/\mu$.
Radiative corrections in this case are less important.

The nontrivial phase structure for $m^2<0$ may be understood
as follows:
We first note that since $\phi_i(y)$ obey the twisted
(antiperiodic) boundary condition, a non-vanishing vacuum
expectation value of $\langle\phi_i(y)\rangle$ immediately 
implies that
it is $y$-dependent, otherwise it would not satisfy the boundary
condition.
The $y$-dependent configuration of $\langle\phi_i(y)\rangle$ 
will induce the
kinetic energy proportional to $1/R^2$.
It follows that for large radius $R$, non-vanishing 
$\langle\phi_i(y)\rangle^2$
is preferable because the origin is not the minimum of
the potential for $m^2<0$
and because the contribution from the kinetic energy
is expected to be small.
Therefore, for large radius $R$, the $O(N)$ symmetry and also the 
translational invariance of $S^1$ are spontaneously broken.
Since $\langle\phi_i(y)\rangle$ must obey the antiperiodic 
boundary condition, non-vanishing $\langle\phi_i(y)\rangle$
cannot be constants and turn out to be of the form
$\langle\phi_i(y)\rangle=(v\cos(y/2R),v\sin(y/2R),0,\cdots,0)$,
where $v=\sqrt{(2\mu^2-1/(2R^2))/\lambda}$ at the tree level.
Since two of $\langle\phi_i(y)\rangle$'s have non-vanishing 
expectation values, the $O(N)$ symmetry should be broken to $O(N-2)$,
but not $O(N-1)$\footnote{
The exception is the model with $N=1$.
In this case, there is no continuous symmetry and the $O(1)$
model has only a discrete symmetry $Z_2$.
The vacuum expectation value of $\phi(y)$ is found to be
a kink-like configuration for $R>R^\ast$\cite{translation}.
}.
On the other hand, for small radius, the contribution from
the kinetic energy becomes large, so that the $y$-independent
configuration of $\langle\phi_i(y)\rangle$ is preferable 
and this implies
that $\langle\phi_i(y)\rangle$ should vanish.
\section*{(3) 
$
U=
\left(
\begin{array}{cc}
  {\bf 1}_L& 0     \\
  0                              & -{\bf 1}_{N-L}     
\end{array}
\right)
$
}

Since the twist matrix $U$ is not proportional to
the identity matrix,
the boundary condition (\ref{2})
explicitly breaks $O(N)$ down to
$O(L)\times O(N-L)$,
which is the subgroup of $O(N)$ commuting with $U$.
For $m^2>0$,
the $O(L)\times O(N-L)$ symmetry is unbroken in
a whole range of $R$ if $N>L>(N-4)/3$,
but is broken to $O(L-1)\times O(N-L)$ for
$R<R^\ast={\cal O}(\sqrt{\lambda}/m$)
if $0<L<(N-4)/3$,
in spite of positive $m^2$.
This symmetry breaking for $R<R^\ast$ comes from
the fact that a one-loop self-energy diagram
in which a boson obeying the antiperiodic boundary condition
propagates gives a
negative contribution to the squared mass\cite{radiative2}.

For $m^2<0$,
the $O(L)\times O(N-L)$ symmetry is broken to
$O(L-1)\times O(N-L)$ in a whole range of $R$
if $0<L<(N-4)/3$,
but is restored for $R\leq R^\ast={\cal O}(\sqrt{\lambda}/\mu)$
if $N>L>(N-4)/3$.
It should be noticed that the translational invariance is not
broken in this model because the vacuum expectation values
of the untwisted bosons are always $y$-independent and
because no twisted bosons acquire non-vanishing
vacuum expectation values.
%
%
%
%
%
%\section*{(4) 
%$
%U=
%\left(
%\begin{array}{ccc}
%  r(\alpha)&\cdots&0  \\
%  :&\ddots&: \\
%  0&\cdots&r(\alpha)     
%\end{array}
%\right),\\
%\quad\quad r(\alpha)\equiv
%\left(
%\begin{array}{cc}
%  \cos(2\pi\alpha)&-\sin(2\pi\alpha)\\
%  \sin(2\pi\alpha)&\cos(2\pi\alpha)
%\end{array}
%\right)
%$
%}
%
%
%
%In this case, the boundary condition (\ref{2}) turns out
%to explicitly break the $O(N)$ symmetry to $U(N/2)$,
%which is the subgroup of $O(N)$ commuting with the
%twist matrix $U$.
%For $m^2>0$, the $U(N/2)$ symmetry remains unbroken
%in a whole range of $R$.
%For $m^2<0$ and $R>R^\ast\sim\alpha/\mu$, the $U(N/2)$
%symmetry is spontaneously broken to $U(N/2-1)$ and
%the translational invariance of $S^1$ is also broken
%spontaneously.
%For $R\leq R^\ast$, all the broken symmetries are restored.
%This symmetry restoration for $R\leq R^\ast$ has a classical
%origin, as in the case of $U=-{\bf 1}$.
%
%
%
%
%
\section*{(4) General $U\in O(N)$}

We can show that the twisted boundary condition (\ref{2})
generally breaks $O(N)$ to 
$O(L_0)\times U(L_1/2)\times \cdots \times U(L_{M-1}/2)\times O(L_M)$
with $L_0+L_1+\cdots +L_M=N$\cite{o(n)}.
A new phenomenon is that in some class of models phase transitions
could occur several times when the radius $R$ varies from $0$ 
to $\infty$.
The full details of the phase structure will be reported 
elsewhere\cite{o(n)2}.
\section*{Acknowledgments}
We would like to thank to J. Arafune, C.S. Lim,
M. Tachibana and K. Takenaga for valuable discussions
and useful comments.
This work was supported in part 
by JSPS Research Fellowship for Young Scientists (H.H)
and by Grant-In-Aid for 
Scientific Research (No.12640275)
from the Ministry of Education,
Science, and Culture, Japan (M.S).
%\begin{appendix}
%%\section*{Appendix}
%We can insert an appendix here and place equations so that they are
%given numbers such as Eq.~(\ref{eq:app}). 
%\begin{equation}
%x = y.
%\label{eq:app}
%\end{equation}
%\end{appendix}

\end{document}